\documentclass[prl,aps,a4paper,twocolumn,superscriptaddress,nofootinbib]{revtex4-2}

\usepackage[T1]{fontenc}
\usepackage[utf8]{inputenc}
\usepackage{amssymb}
\usepackage{amsmath}
\usepackage{graphicx}
\usepackage{bbm}
\usepackage{psfrag}
\usepackage{hyperref}
\usepackage[normalem]{ulem}

\usepackage{xcolor}
\hypersetup{
    colorlinks=true,
    citecolor=blue,
    linkcolor=blue,
    filecolor=magenta,
    urlcolor=blue}
\definecolor{ccqqqq}{rgb}{1,0,0}
\definecolor{uuuuuu}{rgb}{0.26666666666666666,0.26666666666666666,0.26666666666666666}
\definecolor{qqwwzz}{rgb}{0,0,1}
\usepackage{float}

\newcommand{\beq}{\begin{equation}}
\newcommand{\eeq}{\end{equation}}
\newcommand{\bea}{\begin{eqnarray}}
\newcommand{\eea}{\end{eqnarray}}
\newcommand{\bit}{\begin{itemize}}
\newcommand{\eit}{\end{itemize}}

\def\p{\partial}

\usepackage{geometry}
\newgeometry{includefoot,left=2cm,right=2cm,bottom=1cm,top=2cm}

\newcommand{\eg}{{\it e.g.,}\ }
\newcommand{\ie}{{\it i.e.,}\ }

\begin{document}

\author{Constantin Bachas}\email{costas.bachas@ens.fr}

\affiliation{\it Laboratoire de Physique  de l'\'Ecole Normale Sup\'eri{e}ure, \\
CNRS, PSL  Research University  and Sorbonne Universit\'es,
24 rue Lhomond, 75005 Paris, France}

\author{Stefano Baiguera}\email{baiguera@post.bgu.ac.il}

\affiliation{\it Department of Physics, Ben-Gurion University of the Negev, \\ David Ben Gurion Boulevard 1, Beer Sheva 84105, Israel}

\author{Shira Chapman}\email{schapman@bgu.ac.il}  

\affiliation{\it Department of Physics, Ben-Gurion University of the Negev, \\ David Ben Gurion Boulevard 1, Beer Sheva 84105, Israel}

\author{Giuseppe Policastro}\email{giuseppe.policastro@ens.fr}

\affiliation{\it Laboratoire de Physique  de l'\'Ecole Normale Sup\'eri{e}ure, \\
CNRS, PSL  Research University  and Sorbonne Universit\'es,
24 rue Lhomond, 75005 Paris, France}

\author{Tal Schwartzman}\email{taljios@gmail.com\\ \\}

\affiliation{\it Department of Physics, Ben-Gurion University of the Negev, \\ David Ben Gurion Boulevard 1, Beer Sheva 84105, Israel}

\title{Energy Transport for Thick Holographic Branes}
\begin{abstract}
Universal properties of two-dimensional conformal  interfaces are encoded by the flux of energy transmitted and reflected during a scattering process.
We develop an innovative  method that allows us to use  results for the energy transmission in thin-brane holographic models to find the energy transmission for general smooth domain-wall solutions of three-dimensional  gravity. Our method is based on treating the continuous geometry as a discrete set of branes. 
 As an application, we  compute the transmission coefficient of a  Janus interface in terms of its deformation parameter.
\end{abstract}

\maketitle

\noindent \emph{1. Introduction and summary.--} Defects and interfaces\\ are important probes in  
 quantum field theory  and  also ubiquitous in condensed-matter physics. They include such diverse systems as
   junctions of quantum wires, constrictions  of quantum Hall liquids,  or impurities in quantum spin chains and ultracold atomic gases; see, 
 \eg \,\cite{PhysRevLett.68.1220,PhysRevB.46.10866,Wong:1994np,PhysRevLett.74.3005,
 Oshikawa:1996dj,PhysRevB.59.15694,PhysRevLett.82.402,PhysRevB.62.4370,PhysRevA.85.023623,PhysRevB.94.085116,PhysRevLett.121.026805}.  
 Many of these systems exhibit physical phenomena that are not present  when the theory is weakly coupled, and it is important to  develop  analytic tools   to study them.
 In this Letter, we use the AdS/CFT correspondence  \cite{Aharony:1999ti}, which provides a dictionary between strongly-coupled field theories and gravitational systems, to calculate  the transmission of energy through a large class of  interfaces with holographic duals.

We focus on conformal  interfaces between two 2D conformal  field theories, $\mathrm{CFT}_L$ and $\mathrm{CFT}_R$. 
Folding along the interface gives a  theory living on  half-space   with two  energy-momentum  tensors,  $T_L$ and $T_R,$ which are separately conserved in the bulk. 
Their sum $T_{\rm tot} = T_L + T_R$ is also conserved at the interface,  but this is not true in general  for the other spin-2 current,   $T_{\rm rel} = c_RT_L - c_LT_R$ 
 where $c_L, c_R$ are the central charges of the two CFTs.  
As shown in Ref.\, \cite{Meineri:2019ycm} the fraction of energy transported across such 2D interfaces   
is universal, \ie 
the same for all incident excitations,  and controlled by a parameter $c_{LR}$ that appears  in the two-point function of $T_{\rm rel}$ \cite{Quella:2006de,Billo:2016cpy}. 
 Explicitly, 
 \beq\label{first} 
 c_{LR} = c_L {\cal T}_L = c_R {\cal T}_R 
 \eeq
 with ${\cal T}_L\, ({\cal T}_R)$ the fraction of transmitted  energy for  excitations hitting the interface from the left\,(right). 
 Note that the second equality in \eqref{first} is   the  condition for detailed balance.
  Together with the ground-state entropy $\log g$ \cite{Affleck:1991tk},    $c_{LR}$ is an important
 parameter of  any  2D interface CFT (ICFT).

Holographic studies of energy transport   \cite{Bachas:2020yxv,Baig:2022cnb}  were  so far limited to  toy models involving thin branes anchored on the  
boundary of anti-de Sitter (AdS) spacetime  \cite{Karch:2000ct,Karch:2001cw,DeWolfe:2001pq,Bachas:2001vj}. In this Letter, we will move from these ad-hoc models to a whole new arena of holographic interfaces described by  
smooth (super)gravity solutions, which we refer to as thick branes; see, \eg \cite{Bak:2003jk,Freedman:2003ax,Bak:2007jm,Bak:2011ga,Chiodaroli:2009yw,Chen:2020efh,Lozano:2021rmk}.
Such solutions have well-defined, rather than hypothetical, dual ICFTs,  thus  placing the analysis on much firmer ground.

While the entanglement entropy, complexity and other quantum-information measures were extensively studied in holography for both thin-brane models and for smooth  solutions \cite{Azeyanagi:2007qj,Takayanagi:2011zk,Estes:2014hka,DHoker:2014qtw,Erdmenger:2015spo,Gutperle:2015hcv,Gutperle:2016gfe,Karch:2021qhd,Anous:2022wqh,Karch:2022vot,Chapman:2018bqj,Braccia:2019xxi,Sato:2019kik,Hernandez:2020nem,Auzzi:2021nrj,Baiguera:2021cba,Auzzi:2021ozb}, less attention has been paid   to energy transport and to the parameter  $c_{LR}$. 
For a single thin brane separating two patches of AdS$_3$ this parameter was   first computed  in ref.\,\cite{Bachas:2020yxv} with the result
\beq\label{cLR}
c_{LR} = {3\over G}   \left( \frac{1}{\ell_L} + \frac{1}{\ell_R} + 8 \pi G\, \sigma  \right)^{-1} \, , 
\eeq
where $\sigma$ is the brane tension, $G$ is the three-dimensional Newton constant,  and 
$\ell_L, \,\ell_R$
are the 
asymptotic AdS$_3$ radii,   related to the central charges by the Brown-Henneaux formula  \cite{Brown:1986nw}   $c = 3 \ell  /2G$. 
The above result  was  confirmed in  Ref.\,\cite{Bachas:2021tnp} from  nonequilibrium steady states of the thin-brane  model.
 Our goal here is to compute $c_{LR}$  for general two-dimensional ICFTs that admit a smooth holographic dual of 
three-dimensional  gravity coupled to any kind of matter. 

We bypass the computationally involved nature of the problem by treating the geometry as a  discrete set of thin-branes. 
Our argument is inspired by the observation \cite{Baig:2022cnb}  (see also \cite{Bachas:2021tnp})
that  transmission past  a pair of   interfaces modeled by thin branes with tensions $\sigma_1, \sigma_2$
  is the same as for a single thin brane with tension $\sigma_1+ \sigma_2$. That is, the transmission is additive in the tensions.

The SO(2,1) symmetry of the ICFT ground state implies that the dual geometry is a  warped product  $\mathbb{R}\times_w$AdS$_2$, and that all matter-field backgrounds 
 only depend on  the proper-distance  coordinate $y$ on $\mathbb{R}$. It follows that the matter energy-momentum tensor can be written as
   \begin{equation}\label{Tmn}
  T_{\mu\nu}^{\rm matter} =  -\Lambda \,g_{\mu\nu} -  {d\sigma\over dy}   \,\Pi_{\mu\nu}\ , 
  \end{equation}
  where $\Pi_{\mu\nu}$ is a projector on  AdS$_2$ and $\Lambda$ and  $d\sigma/dy$ are functions only of $y$.
  They are the vacuum energy and  tension density of our brane array. We must assume that $\Lambda(y)$ is everywhere negative, so that between two branes the geometry is always AdS. 
We argue that  for such solutions $c_{LR} $ is still given by Eq.\,\eqref{cLR} with the replacements
  \begin{equation}\label{maini}
  \sigma \to \int_{-\infty}^\infty  {d\sigma\over dy}\,dy  \quad {\rm and} \ \  \begin{cases} \ell_{L} \to 
  1/\sqrt{-\Lambda (\mbox{\small $-\infty$}) } \,,
   \\ %[.4em]  
   \ell_{R} \to 1/\sqrt{-\Lambda(\mbox{\small $\infty$}) }
  \ . 
  \end{cases}
  \end{equation}
This  follows from iterating the additivity argument of  \cite{Baig:2022cnb} and approximating \eqref{Tmn}  by  a  dense array of thin branes 
with  tensions  $d\sigma$.
The above simple formula for  $c_{LR}$ in terms of the  dual geometry  is the main result of the Letter.

We  verified, using as matter a scalar field,  that the 
 above result  obeys the ANEC bounds \cite{Meineri:2019ycm}
 \begin{equation}\label{anec1}
 0\leq c_{LR} \leq {\rm min} (c_L, c_R) \ , 
  \end{equation} 
\ie the condition that both the reflected and the transmitted energies are always positive. We further show that, in contrast to the thin-brane model, these bounds can be  saturated. 
Total reflection, in particular, is possible without depleting  the  degrees of freedom of one side, as in Ref.\,\cite{Bachas:2020yxv}.

A key part of the argument leading to the result \eqref{cLR}-\eqref{maini}  is that    $c_{LR}$
 can be extracted from transverse-traceless graviton modes in AdS$_2$, whose linearized wave equation  depends  only on  the geometry, not on the  matter-field backgrounds \cite{Csaki:2000fc,Bachas:2011xa}.
Since the  thin-brane discretization reproduces (by construction, as we will see)  the smooth background geometry in the continuum   limit,  
  it should  also give the same linear  scattering of transverse-traceless  modes.

The results of this Letter could be of use to a large fraction of the physics community, both condensed-matter theorists modeling strongly-coupled interfaces and defects and those interested in quantum gravity and holography. The relation between entanglement and energy transfer is, for example, important in  recent calculations of quantum extremal surfaces and islands,  where both thin-brane toy models and thick branes have been  employed \cite{Almheiri:2019psf,Penington:2019npb,Almheiri:2019qdq,Almheiri:2020cfm,Bak:2020enw,Uhlemann:2021nhu,Demulder:2022aij,Karch:2022rvr}. On a different note,  a double Wick rotation  converts our thin-brane array to a  cosmology of discrete quenches at which the
 Hubble parameter jumps. It could be interesting to explore  if such a discretization can help  calculate  the cosmological production and  evolution of gravity waves.

\vspace{10pt}
\noindent \emph{2. AdS domain walls.--}
Consider Einstein gravity coupled to a  scalar field -- the extension  to arbitrary matter  will be straightforward.
 The action, in units $8\pi G=1$, reads
\beq 
I_{\rm gr} = \frac{1}{2} \int  d^{n+1}x \sqrt{-g} \,\,  [  R   -   \p^{\mu} \phi \p_{\mu} \phi  - 2 V(\phi) ]  \, . 
\label{eq:initial_Janus_action}
\eeq
The dimension $n+1$ is arbitrary for now. We later  set $n=2$.
Solutions dual to vacuum ICFTs depend on a space coordinate $y \equiv x^n$ such that (with $\alpha, \beta = 0, \dots, n-1$)
\beq\label{propertime}
\phi = \phi (y)\, ,  \ \ ds^2 =   dy^2 +   a^2(y)\, \bar\gamma_{\alpha\beta} dx^\alpha dx^\beta\ , 
\eeq
and $\bar \gamma_{\alpha\beta}$ is the metric of unit-radius AdS$_n$.
It is convenient to define the conformal coordinate $\theta$,   
such that  $a\, d \theta = dy $ and
 \beq\label{thetam}
  \ \ ds^2 =  a^2(\theta) \bigl(\, d\theta^2 + \bar\gamma_{\alpha\beta} dx^\alpha dx^\beta\bigr) \ .  
\eeq
The field $\phi(y)$ and  scale factor $a(y)$ obey 
\begin{equation}\label{FRW1}
{n(n-1)\over 2} \left[ \left({a^\prime\over a}\right)^2- {k\over a^2} \right] = {1\over 2}  ({\phi^\prime })^2 - V(\phi)\  
\end{equation}
 \vskip -5mm
\begin{equation}\label{FRW2}
{\rm and}\  \ \ \ {\phi^{\prime\prime} }  + n  \phi^\prime\,{a^{\prime} \over a}  -   {dV\over d\phi} = 0\ , 
\end{equation}
where   $k=-1$ is the curvature of $\bar\gamma_{\alpha\beta}$, primes denote derivatives with respect to $y$, and $V(\phi)$ is the potential. 
At a critical point $\phi_c$ of $V$, the solution 
 \beq\label{eq:trivial}
 \phi =\phi_c\ \ \  {\rm and}  \quad  a  = {\ell\over \cos \theta} = \ell\,\cosh ({y/ \ell}) \  
 \eeq
describes pure  AdS$_{n+1}$ spacetime
of radius $\ell$\,,  where $V(\phi_c) = -n(n-1)/2\ell^2$\,.  
 Domain-wall (DW) solutions interpolate  between  critical points.

A double Wick rotation,  $y \to i t$ and
AdS$_n\to$EAdS$_n$ converts Eqs.\,\eqref{FRW1} and \eqref{FRW2} to the familiar  equations for  an open,  homogeneous and  isotropic  universe coupled to  an inflaton $\phi$ \cite{Mukhanov:2005sc}.
Another  related  context \cite{Boonstra:1998mp,Girardello:1998pd,Freedman:1999gp}, 
with $k=0$, describes holographic renormalization-group flows triggered by an operator dual  to $\phi$. 
  
\begin{figure}[htb]
\centering
\includegraphics[scale=0.14]{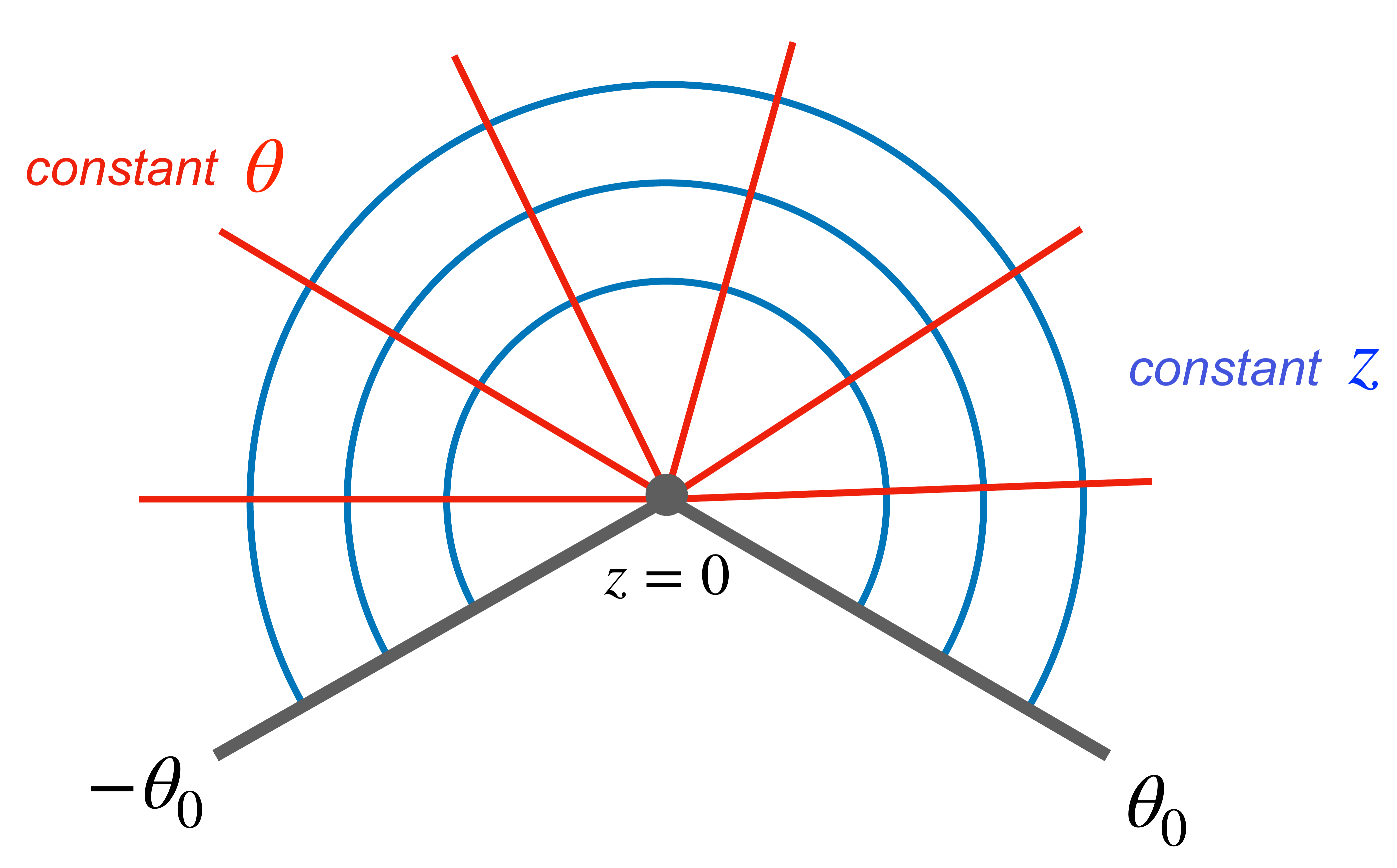}
%\vskip -12mm
\caption{Foliation of the Janus geometry by $\mathrm{AdS}_n$ fibers with radial Poincaré coordinate    $z$. 
The dual  CFT interface is located at $z=0$, whereas the conformal boundary also includes $\theta = \pm \theta_0$  where $a(\theta)$ diverges. }
\label{Janusfig}
\end{figure}

Here we are interested in AdS domain walls with metric  given by  \eqref{thetam}. 
Figure~\ref{Janusfig} shows a typical geometry parametrized by $(z, \theta)$  with
  $z\equiv x^{1}$  the radial Poincar\'e coordinate of AdS$_n$ (\ie $\bar\gamma_{\alpha\beta} = \eta_{\alpha\beta}/z^2$).  
For  pure AdS$_{n+1}$,   $(z, \theta)$ are  polar coordinates on the  $(\xi, u)$ plane,  with $\xi$  
the radial  Poincar\'e  coordinate of AdS$_{n+1}$ and $u$ the coordinate transverse to the would-be  interface.
More generally  $\theta$ takes values in an interval $(-\theta_0, \theta_0)$, the field
  $\phi(\theta)$  approaches critical points $\phi_{L}$ and $\phi_R$ of $V$ at the two extremes,  and
 \beq\label{Asradii}
a\bigl(-\theta_0 +\delta\theta\bigr)   \simeq {\ell_{L}  \over  \delta \theta}\,, \quad
a\bigl(\theta_0 -\delta\theta\bigr)   \simeq {\ell_{R}  \over  \delta \theta} 
\eeq
with $\ell_{L}$,  $\ell_{R}$   the asymptotic AdS$_{n+1}$ radii.

\smallskip 
A  simple example of holographic ICFT 
for $n=2$ is the Janus solution found in \cite{Freedman:2003ax,Bak:2007jm}.
In this case the asymptotic radii are $\ell_L=\ell_R \equiv L$ and $\phi$ is the dilaton  that is dual to a marginal  operator,  so that $V(\phi)\equiv -1/L^{\,2}$ is constant. 
Equation \,\eqref{FRW2} then shows that $\phi^\prime a^2$ is also constant. Inserting in Eq.\,\eqref{FRW1} 
 gives   
\begin{subequations}\label{13}
\begin{align}
\label{eq:profile_f}
a &= \frac{L}{\sqrt{2}} \left[ 1 + (1-b) \cosh ({2y/ L})\right]^{1/2}   \, , \\
\phi  &= {\phi_0 } + {1\over \sqrt{2}} \log\left[  \frac{\sqrt{2-b} + \sqrt{b} \,  \tanh (y/L)}{\sqrt{2-b} - \sqrt{b} \, \tanh (y/L)} \right]  \, , 
\label{eq:dilaton}
\end{align}
\end{subequations} 
where 0\,$\leq$\,$b$\,$\leq$\,1.\footnote{
$b  \equiv 1-\sqrt{1-2\gamma^2}$ comparing with the conventions of \cite{Bak:2007jm}.}
As before, we can trade $y$ for $\theta$ satisfying $d\theta=dy/a$.
As $y$ ranges from $-\infty$ to $\infty$, 
$\theta$ varies from $-\theta_0$ to $\theta_0$ where 
\beq
\theta_0= \left(1  - {b\over 2}\right)^{-1/2} K\left({b \over  2-b }\right)    \geq \pi/2
\eeq
with $K$ the complete  elliptic integral of the first kind. 
For  $b=0$ the dilaton is constant and the geometry is pure AdS$_3$,  whereas for  $b=1$, the dilaton is linear
and the geometry is $\mathbb{R}\times$AdS$_2$.  More generally the arbitrary parameters $\phi_0$ and $b$ can be traded for  $\phi(\pm\theta_0)$, 
  the marginal couplings of the CFT on the two sides of the interface.
 
 The Janus solution \eqref{eq:profile_f} and \eqref{eq:dilaton} can be embedded in type IIB supergravity compactified on $\mathrm{AdS}_3 \times S^3 \times  T^4$ 
or $\mathrm{AdS}_3 \times S^3 \times K3$,  but it is nonsupersymmetric and  possibly unstable.  
In addition to activating more fields,  
supersymmetric solutions have  
a nontrivial dependence on the extra dimensions 
\cite{Chiodaroli:2009yw,Chen:2020efh,Lozano:2021rmk}.
  Our discussion below will \textit{a priori} apply only after  this dependence has been  smeared,  or   better if the reduction to $n+1$ dimensions is a consistent truncation.

\begin{figure}[ht]
\centering
\includegraphics[scale=0.6]{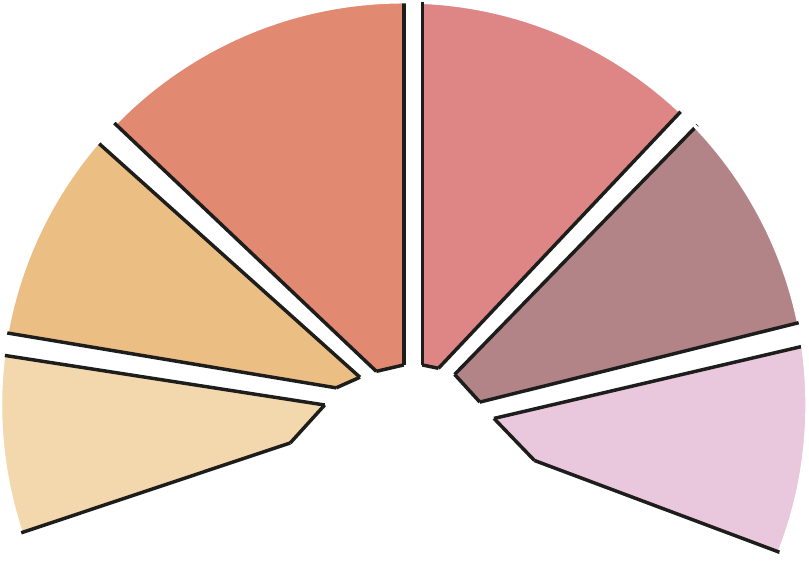}
%\vskip -9mm
\caption{$N_\epsilon$  regions of AdS$_{n+1}$ spacetime glued together along thin branes. 
 In the continuum limit $\epsilon\to 0$, this geometry will match the smooth  solution.}  
 \label{fig-discrete_geometry}
\end{figure}

 \vspace{10pt}
\noindent \emph{3. Matter as a thin-brane array.--} 
 Now the key idea  is to replace the smooth solutions by a ``pizza'' of  AdS$_n$ slices separated by thin  tensile branes; see Fig.~\ref{fig-discrete_geometry}.
   We divide  the range of $\theta$ into $2N_\epsilon$ intervals
of size $\epsilon$ (so that $\epsilon N_\epsilon = \theta_0$),   and  let the metric in the $j$-th   interval  
 be AdS$_{n+1}$  with radius $\ell_j$. The  label $j$ runs from $-N_\epsilon+1$ to $N_\epsilon$. 
 In the $j$th interval 
 the modified  scale factor, $\tilde a $,   reads
  \beq\label{iter1}
 \tilde a(\theta) = {\ell_j\over \cos (\theta - \delta_j)}\quad {\rm for} \ \ 
  (j-1)\epsilon  <\theta < j \epsilon\, .
 \eeq
Furthermore the two end point values of each interval are fixed to be those of  the original solution, $\tilde a (j\epsilon) = a(j\epsilon) \equiv a_j $ for all $j$. 

Using Eq.\,\eqref{iter1} we can  express  $\ell_j$ and  the shift $\delta_j$  in terms of the radii of the AdS$_n$ 
branes  $a_j$ and $a_{j-1}$. 
 The $\ell_j$ will \textit{a priori} jump from one interval to the next, and we attribute this jump  to thin-brane sources with tension   $\sigma_j$,  localized at  $\theta=j\epsilon$. 
The Israel  junction conditions \cite{israel1966singular}
read (see, \eg \cite{Bachas:2021fqo})
\beq\label{iter2}
\sigma_j a_j =   \sqrt{\left({a_j\over \ell_j}\right)^2 - 1}\, - \, \sqrt{\left({a_j\over \ell_{j+1}}\right)^2 - 1} \ , 
\eeq
for $\ell_{j+1} > \ell_j$ (otherwise the sign of the rhs should be flipped to maintain positive tension).
Equation  \eqref{iter2}  expresses $\sigma_j$ in terms of $a_j$, $\ell_j$ and $\ell_{j+1}$ or, using \eqref{iter1}, in terms of the three scale factors $a_{j-1}, a_j$, and $a_{j+1}$.

     By construction,  the   discretized (tilde) geometry  approaches  the smooth ICFT solution  as 
    $\epsilon\to 0$. Both  $\ell_j$ and $\delta_j$ approach smooth functions  $\ell(\theta)$ and $\delta(\theta)$ in the limit, while
    $\sigma_j/\epsilon$ tends (as a distribution)
    to a  tension density $d\sigma/d\theta$.
   From  Eqs.\,\eqref{iter1} and \eqref{iter2} 
\begin{subequations}\label{sigma1}
\begin{align} 
\ell =& {a\over \sqrt{1+ (\dot a /a)^2}}\ , \quad 
\tan(\theta - \delta) = {\dot a  \over a}\  \\
&\, \ \ \ {\rm and} \qquad  {d\sigma\over  d\theta} = {a \,\vert\, \dot \ell \,\vert  \over \ell^2 \sqrt{a^2 - \ell^2}}\ , 
\end{align}
\end{subequations} 
where dots  stand for derivatives  with respect to $\theta$.

Since the  discretized geometry obeys the Einstein equations, 
its  source  should  also
converge to the  energy-momentum  tensor of the  smooth solution.
For a single scalar field, we have 
 \beq
\begin{aligned}
 T_{\mu\nu}^\phi  
 = -   (\partial^\rho \phi \partial_\rho \phi )\,  {\Pi}_{\mu\nu} + g_{\mu\nu}   \left( {1\over 2} \partial^\rho \phi \partial_\rho \phi - V\right)  \, , 
\end{aligned}
\eeq
where    ${\Pi}_{\mu\nu} $=$g_{\mu\nu}-$ $\,\hat n_\mu \hat n_\nu$ projects onto  hypersurfaces   of   constant $\phi$
(with $\hat n_\mu$  the
unit normal   vector field). 
 Comparing to the thin-brane array,   which includes a piecewise-constant vacuum energy
  $\Lambda = -1/\ell^{2}$,  
 we get
\beq\label{sigma2}
  \Lambda =   -{1\over 2}   (\phi^\prime)^2 + V   \ \  \ 
  {\rm and} \quad \   d  \sigma  \, =\,  (\phi^\prime)^2 \,  dy  \ . 
\eeq
For the second equality we  used the fact  
that the   energy-momentum  tensor of  a thin brane of tension $\lambda$
localized at $y=y_0$ is 
$T_{\mu\nu} = - \lambda  \Pi_{\mu\nu}\, \delta(y-y_0)$ (see, \eg \cite{Shiromizu:1999wj}). 
Using  Eqs.\,\eqref{FRW1} and \eqref{FRW2}, one can check  that the expressions \eqref{sigma1} and \eqref{sigma2} for $\Lambda$ and $d\sigma$  
indeed coincide.  
 For the Janus solution \,\eqref{13}, we find
  \beq
 d \sigma^{\rm Jan}   = {2b(2-b) \, dy \over L^2 [1 + (1-b) \cosh(2y/L)]^2}\ \,. 
  \eeq
Integrating gives the total tension of the brane array 
\beq\label{totJanus}
\sigma_{\rm tot}^{\rm Jan} = {4\over L \sqrt{b(2-b})}\, {\rm arctanh}\Bigl( \sqrt{b\over 2-b}\,\Bigr) - {2\over L}\ .
\eeq
This vanishes for $b=0$, as it should,  and diverges in the opposite limit $b\to 1$.

In approximating the energy-momentum tensor by a thin-brane array, we only relied on the AdS$_n$ isometries guaranteeing that $T_{\mu\nu}^{\rm matter}$ is a linear combination of $g_{\mu\nu}$ and  $\hat n_\mu \hat n_\nu$, see Eq.~\eqref{Tmn}.
Simple algebra leads then to a formula  valid in any dimension with any  matter content, 
 \beq\label{tensionu}
d \sigma =   \Bigl( T_{yy}^{\rm matter} - {1\over n} \Pi^{\mu\nu}\,T_{\mu\nu}^{\rm matter} \Bigr)\, dy\ .  
 \eeq
 We now focus on $n=2$ and explain why the above tension density controls the energy transport.

\vspace{10pt}
\noindent \emph{4. Scattering off a thin-brane array.--}
The   discretized geometry makes it possible to exploit the results of Refs.\,\,\cite{Bachas:2020yxv,Bachas:2021tnp,Baig:2022cnb}
on energy transport in thin-brane  models. For an isolated  brane, the classical linearized  scattering calculation gives  \cite{Bachas:2020yxv} 
\beq\label{tb1}
\mathcal{T}_{1\to 2} = \frac{2}{\ell_{1}} \left(\, \frac{1}{\ell_1} + \frac{1}{\ell_2} +  8\pi G\, \sigma  \right)^{-1} \, ,
 \eeq 
where $\mathcal{T}_{1\to 2}$ is the fraction of energy incident from side 1 and transmitted to side 2 of the  ICFT,
$\ell_{1,2}$ are the AdS radii,  $\sigma$ is the brane tension, and we have restored the $8\pi G$. 
 Multiplying by $c_1$ recovers Eq.\,\eqref{cLR}. 

\begin{figure}[ht] 
\centering
\hspace*{-0.5cm}
\includegraphics[scale=0.12]{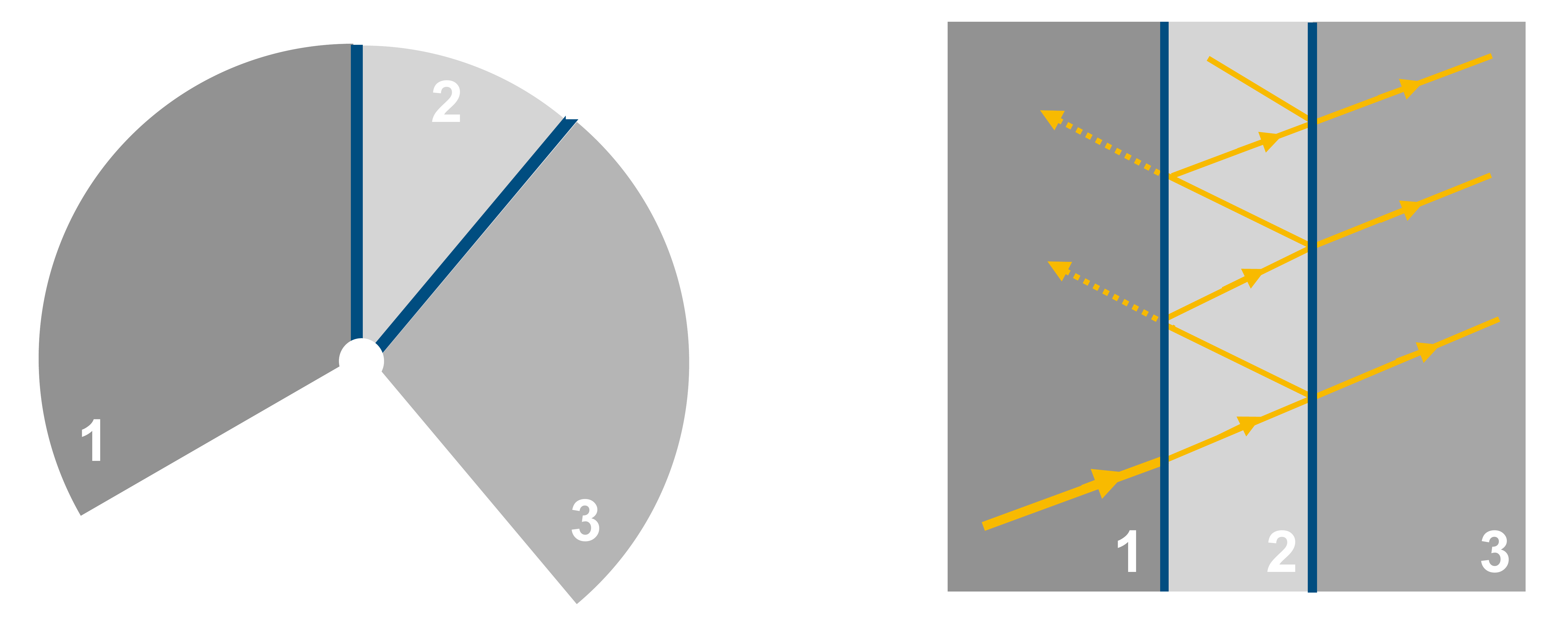} 
%\vskip -16mm
\caption{Fixed-time slice of the  brane-pair  geometry (left), and the dual boundary ICFT with time running upward (right). 
 The total transmitted energy is  the sum
over the number, $m$, of double reflections in  region 2. The interfaces were infinitesimally separated. }
\label{2branes}
\end{figure}

 This calculation was extended in Ref.\,\cite{Baig:2022cnb} to a pair of thin branes with tensions $\sigma_{1,2}$,  separating  three AdS  regions with
radii $\ell_{1,2,3}$, see Fig.~\ref{2branes}. The result  is that the tensions  add up,
 \beq\label{tb2}
\mathcal{T}_{1\to 3} = \frac{2}{\ell_{1}} \left( \,\frac{1}{\ell_1} + \frac{1}{\ell_3} +  8\pi G( \sigma_1 + \sigma_2) \right)^{-1} \, .
 \eeq
  Both Eqs.\,\eqref{tb1} and \eqref{tb2} can  be also   derived from  nonequilibrium steady states if thermostats  
 are attached on either side of the interface
 \cite{Bachas:2021tnp}. The transport coefficients are in this case extracted by matching the ICFT formula \cite{Bernard:2014qia,Bernard:2016nci} 
\beq
c_{LR}  =  { 12 \, J_E \over \pi (\Theta_L^2 - \Theta_R^2)}\,  , 
\eeq
where $J_E$ is the steady  flow of heat, and $\Theta_{L},\Theta_R$ are the temperatures on the left and right of the interface.
  This  calculation in gravity confirms  that
 energy transport in  2D ICFTs is  universal  \cite{Meineri:2019ycm}.

In the ICFT, Eq.\,\eqref{tb2}  can be interpreted as 
summing over multiple reflections in the middle region, see Fig.~\ref{2branes}.
Indeed, using \eqref{tb1}  one can verify that 
 \beq
\begin{aligned}
\mathcal{T}_{1\to 3} = 
\mathcal{T}_{1\to 2}\mathcal{T}_{2\to 3}\Bigl[ &1 + \mathcal{R}_{2\to 3}\mathcal{R}_{2\to 1}  + \cdots \\
+ &( \mathcal{R}_{2\to 3}\mathcal{R}_{2\to 1})^m + \cdots \Bigr] \ , 
\end{aligned}
\eeq
where $ \mathcal{R}_{1\to 2} = 1 -  \mathcal{T}_{1\to 2}$ and  
 $ \mathcal{R}_{2\to 1}= 1 -  \mathcal{T}_{2\to 1}$. 
 This calculation is classical, not only in gravity but also in   ICFT. 
Quantum interference  can invalidate the   result and lead to 
singular  interface fusion  \cite{Bachas:2007td} which
 signals an instability when the  anchor 
points of the branes get close \cite{Bachas:2021fqo}.   
In the present context we do not expect such instabilities to  arise  if the original (super)gravity solution was stable before the discretization.

Equation\,\eqref{tb2} is easy to extend by iteration to any  number of adjacent  branes. Applying it  to the $\epsilon\to 0$  limit of the thin-brane array leads to
the main result of this Letter,  as summarized by Eqs.\,\eqref{cLR}\,-\,\eqref{maini} in the Introduction.
For the  Janus example in the second section, using the total tension Eq.\,\eqref{totJanus} and $\ell_L=\ell_R = L$  we find
\beq\label{mainJ}
\mathcal{T}^{\rm Jan} = {1\over 2}  \sqrt{b(2-b)} \,\left[ {\rm arctanh}\left(  \sqrt{\frac{b}{2-b}} 
\right)\right]  ^{-1}\ .
\eeq
 Note that  as $b$ varies from $1$ to $0$, $\mathcal{T}^{\rm Jan}(b)$ takes all values in the interval $[0, 1]$.

\vspace{10pt}

\noindent \emph{5. Bounds.--}
As explained in Ref.\,\cite{Bachas:2020yxv}, in the case of a single thin brane the existence of the static vacuum solution requires that the tension 
lies inside the stability window 
\begin{equation}\label{stab}
\left|\frac{1}{\ell_L}-\frac{1}{\ell_R}\right|
\leq 8\pi G \sigma \leq
\frac{1}{\ell_L}+\frac{1}{\ell_R}\ .  
\end{equation}
Inserting these inequalities in Eq.\,\eqref{cLR},  shows that  the ANEC  bounds   \eqref{anec1} are automatically obeyed. 
 As stressed, however, in Ref.\,\cite{Baig:2022cnb}, the upper bound in \eqref{stab} implies (for finite central charges) a positive lower bound on $c_{LR}$,
 so that the thin-brane model does not cover the entire  range allowed by ANEC.  

For the thin-brane array studied here, the  stability window in the continuum limit becomes
\begin{equation}
%-
\left|\frac{d}{dy} \frac{1}{\ell(y)}\right| 
\leq 8\pi G \frac{d\sigma}{dy} \leq
\infty
\end{equation}
The upper bound is trivial because the tension $d\sigma$ is infinitesimal, whereas the local AdS radius is assumed  finite.
 The lower bound is also automatic for any scalar potential, as can be shown using Eqs.~\eqref{sigma2}, the relation $\Lambda(y) = -1/{\ell(y)^2}$, and the equations of motion \eqref{FRW1} and \eqref{FRW2}. 
  Thus the ANEC bounds are also automatic. By relaxing, however, the upper limit of the stability window, one can now cover the entire ANEC range. 
 Note that for Janus  $c_L=c_R$,  and the ANEC bound reduces to $0\leq \mathcal{T}_L= \mathcal{T}_R\leq 1$.

\vspace{10pt}
\noindent \emph{6. Perturbations.--} 
It should be, in principle, possible to compute $c_{LR}$   
 directly for the thick-brane solution.  Without 
 performing  the calculation,    we will  argue that the result is obtained by the  $\epsilon\to 0$ limit of the thin-brane array.

The scattering  states of Ref.\,\cite{Bachas:2020yxv} are characterized by the expectation values  of energy fluxes  incident on or coming out of the interface,
    $\langle T^{\rm in}_L\rangle, \langle T^{\rm in}_{R}\rangle, \langle T^{\rm out}_L \rangle$,  and $\langle T^{\rm out}_{R}\rangle$.  
    The crucial point is that these can be prepared with  metric  perturbations that are transverse and traceless in AdS$_2$, so that their linearized Einstein equations
    depend only on the geometry,  not on the matter-field backgrounds \cite{Csaki:2000fc,Bachas:2011xa}.
 Leaving for the moment the dimension $n+1$
arbitrary,   consider the   perturbation  
  \beq
\begin{aligned}
ds^2 = &dy^2 + a^2(y)\left[ \bar\gamma_{\alpha\beta} + h_{\alpha\beta}\, \right] dx^\alpha dx^\beta\ \\ 
 &
{\rm with} \qquad  \bar\gamma^{\alpha\beta} h_{\alpha\beta} = \bar\nabla^\alpha h_{\alpha\beta}=0\ .   
\end{aligned}
 \eeq
Expanding in AdS$_n$ harmonics, 
\beq
h_{\alpha\beta}({\bf x}, y) = \sum_r h_{\alpha\beta}({\bf x}\vert r)\, \psi_r(y) 
\eeq 
where $r$ labels spin-2 representations of  $\mathrm{SO}(2,n-1)$,
reduces  Einstein's equations to  
\beq\label{KKreduction}
 -(a^n\, \psi_r^\prime)^{\,\prime}\, =\, m^2_r \,a^{n-2} \psi_r\ ,
\eeq  
where $m_r$ is the mass of the excitation in the $r$-th harmonic. For $n>2$ this (Kaluza-Klein) decomposition gives an infinite tower of spin-2 excitations in AdS$_n$
labeled by  $n-1$  momenta and by the mass, or equivalently the scaling dimension  in the dual CFT, 
$m_\Delta^2 = \Delta (\Delta - n + 1)$.  
In the absence of the interface the  spectrum is 
 $\Delta = n, n+1, \dots $.  
This is the decomposition of the massless spin-2 representation of SO(2,$n$) under  SO(2,$n-1$)  \cite{Karch:2000ct,Bachas:2011xa}. 
Note that the  massless  AdS$_{n+1}$ particle gives a tower of massive AdS$_n$ states. 

 The case $n=2$ is special  because the three-dimensional  graviton only has surface excitations. As a result, for a given frequency $\omega$
   the  tower collapses  to a pair of modes, both  of them massless in AdS$_2$ ($\Delta =1$).\footnote{Top-down AdS$_2$ vacua do have  infinite   towers of spin-2 excitations, but these come from the Kaluza-Klein reduction of the extra  seven dimensions, see \eg  \cite{Rigatos:2022ktp}. 
   } 
 They  can be written explicitly as follows
 \beq\label{29}
h_{\pm\pm}({\bf x}, y)\,   =  \,  e^{i\omega (x^0 \pm x^1)}\,\left[\,A^{\omega}_\pm + B^{\omega}_\pm \hskip -1mm \int^y \hskip -1mm {d\tilde y\over a(\tilde y)^2}\,\right],    
 \eeq 
where $A^{\omega}_\pm\,\, B^{\omega}_\pm$ are constants of  integration of Eq.~\eqref{KKreduction}  which is readily solved when $m_r=0$.

  We can use the   constants $A^{\omega}_\pm, B^{\omega}_\pm$  to fix the two ingoing and two outgoing fluxes in the scattering state. 
To this end one must transform to
 Fefferman-Graham (FG) coordinates which are not defined globally,  but must be chosen separately on the   half boundaries  $y=\pm\infty$   (see  
\cite{Papadimitriou:2004rz,Chiodaroli:2010ur,Estes:2014hka,Gutperle:2016gfe}).
Even at leading order in $h_{\alpha\beta}$ these reparametrizations are cumbersome but the main point, following 
\cite{Bachas:2020yxv},  is that the two FG   patches should match across a   static  interface on the AdS$_3$ boundary.  This ensures the conservation of CFT energy,  
  $\langle T^{\rm in}_L\rangle + \langle T^{\rm in}_{R}\rangle = \langle T^{\rm out}_L \rangle+ \langle T^{\rm out}_{R}\rangle$, and eliminates one of the four integration constants.
 
The final equation is a boundary condition at the  Poincar\'e horizon. To obtain it, following again \cite{Bachas:2020yxv}, consider the conjugate ``momenta'' $\pi_{\alpha \beta} = \sqrt{-\tilde g}\, (K_{\alpha \beta} -  \tilde g_{\alpha \beta} K )$, with  $\tilde g_{\alpha \beta}$  the metric on  fixed-$y$ slices and $K_{\alpha \beta}$ the extrinsic curvature.
We can separate $\pi_{\alpha \beta} $ in a traceless part $\hat \pi_{\alpha\beta}$ and the trace. 
At leading order in the perturbation the ``momentum constraint'' actually shows that $\hat \pi_{\alpha\beta}$ is also transverse. Thus, $\hat \pi_{\alpha\beta}$ decomposes into two waves, one going in and one coming out of the Poincar\'e horizon. Setting the outgoing wave to zero,   $\hat \pi_{++}=0$,  gives after a little algebra $B^{\omega}_+=0$. 
This condition, once  imposed at one value of $y$,  remains  valid for all $y$.

The two remaining  free parameters  are used to fix the  incoming energy fluxes in the CFT. The outgoing fluxes
  give then  the  transport coefficients that we set out to compute.  
     The main point about this calculation is that it only  involves the  scale factor $a(y)$, so discretizing it should not change the answer in the continuum limit. Indeed,  a convenient way of  
     performing the calculation is by following exactly the  steps of \cite{Bachas:2020yxv, Baig:2022cnb} for the thin-brane array defined in the third section.

\vspace{10pt}
\noindent \emph{7. Outlook.--} 
Several questions left open by our work deserve further study. 
It would be interesting, in particular,  to extend the calculation of $c_{LR}$   to top-down holographic interfaces that cannot  be reduced to
Einstein's  equations in three dimensions. Another interesting direction is to find a holographic proof of the  fact that the energy-transmission coefficients do not  depend on details of the scattering state \cite{Meineri:2019ycm} or  to understand if our method can be extended to study the holographic transport of electric charge.
Finally, the double Wick rotation mentioned in the second section converts the brane array into a sequence of cosmological quenches between periods  of de Sitter expansion. It would be very interesting to 
understand how our results translate in this context.

\begin{acknowledgments}
\section*{Acknowledgments}
 {We would like to thank  Sara Bonansea, Marc Henneaux, Zohar Komargodski, Yaron Oz and Vassilis Papadopoulos  for useful discussions.
The work of S.B., S.C., and T.S. is supported by the Israel Science Foundation (Grant No. 1417/21) and by the German Research Foundation through a German-Israeli Project Cooperation (DIP) grant “Holography and the Swampland”. 
S.B. is grateful to the Azrieli foundation for the award of an Azrieli fellowship.
S.C. acknowledges the support of Carole and Marcus Weinstein through the BGU Presidential Faculty Recruitment Fund. }
\end{acknowledgments}

\bibliography{biblioJanus}
\bibliographystyle{biblioJanus}

\end{document}